# Heterogeneity of Some Cooperation/Competition Properties[*]

XU Xiu-Lian, FU Chun-Hua, LIU Ai-Fen, HE Da-Ren[**]

*College of Physics Science and Technology, Yangzhou University, Yangzhou 225002*

*We show that the heterogeneity index, which was proposed by Hu and Wang (Physica A 2008 387 3769), can be used to describe the disparity of the cooperation sharing or competition gain distributions that is very important for the cooperation/competition system dynamic understanding. An analytical relation between the distribution parameters and the heterogeneity index is derived, which is in a good agreement with the empirical results. Our theoretical and empirical analyses also show that the relation between the distribution parameters can be analytically derived from so-called Zhang-Chang model (Physica A 2006 360 599 and 2007 383 687). This strongly recommends a possibility to create a general dynamic cooperation/competition model based on Zhang-Chang model.*

Simultaneous cooperation/competition of complex system elements have been a fascinating research topic in commercial, [1-2] ecological, [3] social, [4] and other areas. [5-12] Our research group has been interested in the cooperation/competition systems described by bipartite graphs. [6-12] A bipartite graph contains two types of nodes. One is called "acts", which can be organizations, events, or activities, and the other is called "actors", which are participants. Only different types of nodes in the bipartite graphs can be connected by edges. To describe the collaboration/competition relationship between the actors, a projected unipartite network is often used. In the unipartite network, all the actors, which take part in at least one common act, are connected by equivalent unweighted links. In our previous investigations [6-12] multiple connections were often counted. The collaboration/competition partner number of an actor, *i*, is addressed as its "degree" and is defined as $k_i = \sum_j a_{ij}$.

In the definition $a_{ij}$ represents the "element of the unipartite graph adjacency matrix" and is defined as: $a_{ij} = 1$ if actor *i* and *j* are linked, and $a_{ij} = 0$ otherwise. Correspondingly, there should be two definitions of degrees in the bipartite graph since there are two kinds of nodes. The degree of an actor node *i*, $h_i$, that denotes the act number in which the actor takes part, called "act degree", is defined as $h_i = \sum_j b_{ij}$. In the definition $b_{ij}$ represents the "element of the bipartite graph adjacency matrix" and is defined as: $b_{ij} = 1$ if actor *i* and act *j* are linked, and $b_{ij} = 0$ otherwise. The degree of an act node *j*, $T_j$, that denotes the number of the actors, which take part in the act, called "act size", is defined as $T_j = \sum_i b_{ij}$. Inspired by the models of Barabasi, Albert [13] and Liu et al. [14], Zhang et al. proposed in 2006 an evolution model of the collaboration/competition bipartite networks. [6] In Zhang model, during the networks evolution, the new actors select partners partially randomly (with the probability *p*) and partially by linear preference (with the probability 1-*p*); however, the model for the general cases was not analytically solved. Chang et al. analytically solved the model in 2007. [7] They obtained a conclusion that in the condition, where $T_j$ obeys a unimodal distribution and thus can be approximated as a constant, *T* (Refs. [6] and [7] reported 10 real world networks which show unimodal $T_j$ distributions), both the act degree distribution $P(h)$ and the unipartite graph node degree distribution $P(k)$ (when the multiple links are counted) show the so-called "shifted power law (SPL) functions". The function can be expressed as $P(x) \propto (x+\alpha)^{-\gamma}$. When $\alpha=0$, it takes a power law form. In the condition that $x$ is normalized ($0<x_i<1$ and $\sum_{i=1}^{M} x_i = 1$, $M$ denotes the total number of $x$), we can prove [15] that SPL function tends to an exponential function when $\alpha \to 1$. Therefore an SPL interpolates between a power law and an exponential function, and the parameter $\alpha$ characterizes the degree of departure from a power law. In Zhang-Chang model analysis, $P(h)$ and $P(k)$ show the same $\gamma$ value ($\gamma=1+T/[(T-1)(1-p)]$). For $P(h)$, $\alpha=Tp/(1-p)$; and for $P(k)$, $\alpha=T(T-1)p/(1-p)$. [7] Therefore, it is almost the same to discuss either $P(k)$ or $P(h)$ for a description of the cooperation/competition partner number distribution. In the follows we shall only discuss $P(h)$. Please note that both $h$ and $k$ were not normalized in Refs. [6-7]. If $h$ is normalized as $s=h/(\sum_j h_j)$, the SPL parameter expressions of $s$ distribution will be $\alpha = p/[N(1-p)]$, where $N$ denotes the total number of the actor nodes, and $\gamma$ remains unchanged. The deduction details will not be mentioned due to the page limit.

---
[*] Supported by the National Natural Science Foundation of China under grant Nos 10635040 and 70671089.
[**] To whom Correspondence should be addressed. Email: darendo10@yahoo.com.cn

In addition to the partner number distribution, a more important quantity was proposed to describe the cooperation/competition. [9] Often in each act some actors make concerted effort to accomplish a task and often create a type of production. The production often induces several kinds of resources. The actors, when they are cooperating, are also competing for a larger piece of the resources. For example, some Hollywood actors work together to produce a movie, which should bring ticket office income and famousness. The former is countable, while the later is not. Fu et al. defined a new quantity, "node weight", in 2008. [9] Basically an actor node weight was defined as the part of a countable resource, which the actor shares. However, in a more dissectional consideration, the actual competition intensity of two actors should depend on the act size. Therefore, if $Z_l$ denotes the total countable resource in act $l$, and $z_{il}$ denotes the part shared by actor $i$, the weight of node $i$ in act $l$ was defined as $W_{li}=T_l z_{il}$. The normalized node total weight (NNTW), $\omega_i$, was defined as $\omega_i = (\sum_l T_l z_{il}) / \sum_j [(\sum_l T_l z_{jl})]$. [9-10,15] In Ref. [10] Fu et al. presented the NNTW distribution functions of 14 cooperation/competition networks. All the distributions obeyed SPL, but the parameters, $\alpha$ and $\gamma$, were very different for different systems. Fu et al. showed by empirical data analyses that the values of the parameters indicated the disparity of the actor cooperation shares or competition gains. For this they defined a "disparity in node act weight" as $Y_l = [\sum_{i<j; i,j \in V_l} (W_{li} - W_{lj})^2][(T_l - 1)\sum_{m=1}^{T_l} W_{lm}^2]^{-1}$, where $V_l$ denotes the actor node set of act $l$, and defined the "disparity in node weight of the network" as $Y = (\sum_l Y_l)/N_l$, where $N_l$ denotes the total number of acts. They showed that $Y$ positively and monotonically correlated to both $\alpha$ and $\gamma$, and there was a close relationship, $\gamma \propto \alpha^{0.32}$, between the two parameters. [10] The conclusions are important but may not be accurate due to the lack of the analytic treatment.

There was another way for a description of disparity. To generally quantify degree distribution disparity (which can be viewed as the disparity of the partner number distribution in cooperation/competition networks), Hu and Wang defined a "heterogeneity index", $0 \le H \le 1$, and deduced an analytic expression of $H$ when the degree distribution obeys a power law. [16] As mentioned, power law distribution is an extreme case of SPL.

Actually, both $Y$ and $H$ are pure mathematic descriptions. They should be the same if the discussed quantity distribution functions are in a common form no matter what is the physical meaning of the quantities. Since both the partner number distribution and the NNTW distribution obey SPL in the studied real world cooperation/competition networks, and the analytic expression of $H$ can be deduced for an extreme case, we may use $H$ to describe the heterogeneities of both the distributions if we can extend Hu-Wang analytic expression from power law to SPL. More accurate conclusions should be obtained than what Fu et al. reported in Ref. [10].

Consider $N$ values of a quantity, $u$, which are labeled from 1 through $N$ in increasing order of the $u$ value, i.e., $u_1 \leq u_2 \leq \cdots \leq u_N$. An $x$-$y$ plane is defined where $x_i = i/N$ and $y_i = \sum_{j=1}^{i} u_j$. The disparity of $u$ value distribution can be shown by the $y(x)$ line on the $x$-$y$ plane. As shown in Fig. 1, the heterogeneity index, $H_u$, can be defined as $H_u = S_A/(S_A + S_B) = 1 - 2S_B$,[16] where $S_A$ denotes the area between the diagonal line, $y=x$ (which shows that the distribution is absolutely homogeneous), and the $y(x)$ line; and $S_B$ denotes the area beneath the $y(x)$ line, namely the area between the $y(x)$ line, $y=0$ axis, and $x=1$ axis (the two axes indicate that the distribution is completely heterogeneous since $u$ is concentrated on only one value).

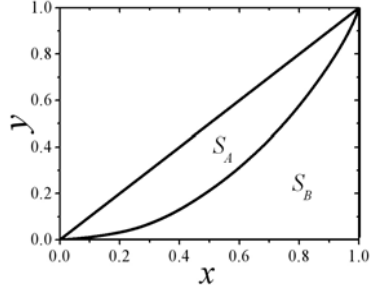

**Fig. 1.** A schematic showing the definition of $H$

For an SPL distribution, such as the cooperation/competition partner number distribution or NNTW distribution, we can deduce the analytic expressions of $H$ following Hu-Wang analytic expression.[16] For NNTW distribution, $P(\omega) \propto (\omega + \alpha)^{-\gamma}$, let $X = \omega + \alpha$, then the $x$-$y$ plane and $H_X$ are defined, respectively, as $x_i = i/N, y_i = (\sum_{j=1}^{i} X_j)/(\sum_{l=1}^{N} X_l)$, and $H_X = 1 - 2S_B = 1 - 2\sum_{i=1}^{N}[(\sum_{j=1}^{i} X_j)/(N\sum_{l=1}^{N} X_l)]$. According to the discussion in Ref. [16] (and the references therein), the rank of $X$ (denoted by $R = 1,2,\cdots,N$) also obeys a power law distribution, i.e., $X(R) = cR^{-\lambda}$, where $\lambda = 1/(\gamma-1)$ and $c$ is a constant. Thus $H_X$ can be written as $H_X = 1 - (2/N)\sum_{i=1}^{N}[(\sum_{j=1}^{i} c\cdot(N+1-j)^{-\lambda})/(\sum_{R=1}^{N} c\cdot R^{-\lambda})]$, $H_{X,\infty} = (2\gamma-3)^{-1}$ (when $\gamma > 2$), or 1 (when $1 < \gamma \leq 2$) ($N \to \infty$) [16].

Since $H_{X,\infty} = 1 - 2\sum_{i=1}^{N} \frac{\sum_{j=1}^{i}(\omega_j + \alpha)}{\sum_{l=1}^{N}(\omega_l + \alpha)} \cdot \frac{1}{N} = 1 - \frac{2\sum_{i=1}^{N}(\sum_{j=1}^{i}\omega_j/N) + N\alpha}{1 + N\alpha}$, and $2\sum_{i=1}^{N}(\sum_{j=1}^{i}\omega_j \cdot \frac{1}{N}) = 1 - \frac{1 + N\alpha}{2\gamma - 3}$ (when $\gamma > 2$), the heterogeneity index of NNTW distribution can be written as $H_\omega \approx H_{\omega,\infty} = (1 + N\alpha)/(2\gamma - 3)$ (when $\gamma > 2$).

Similarly, one has $H_s \approx H_{s,\infty} = (1 + N\alpha)/(2\gamma - 3)$ (when $\gamma > 2$) for the normalized act degree distribution, $P(s) \propto (s + \alpha)^{-\gamma}$. Refer to the aforementioned Zhang-Chang model analytic expressions of $\alpha$ and $\gamma$, we have $H_s = 1/[p + (T+1)/(T-1)]$.

We have presented the definitions of 14 real world collaboration/competition networks and their node weights, the node weight distributions, as well as the SPL distribution parameter values.[15] Now, in order to compare the analytic expression of $H_\omega$ with the empirical investigation results, we show the empirical correlation between $(1 + N\alpha)/(2\gamma - 3)$ and $H_\omega$ of 11 of the systems, which show the $\gamma$ values larger than 2, in Fig. 2. The systems are the 2004 Athens Olympic Game, the university matriculation network, the university independent recruitment network, the undergraduate course selection network of Yangzhou University, the book borrowing network of YZU library, the training institution network, supermarket network, information technique product selling network, notebook PC selling network at Taobao website, Beijing restaurant network, and the human acupuncture point network. For a better specification of the systems, please refer to Ref. [15]. Figure 2 shows that the empirical results are in a rather good agreement with the analytic conclusion.

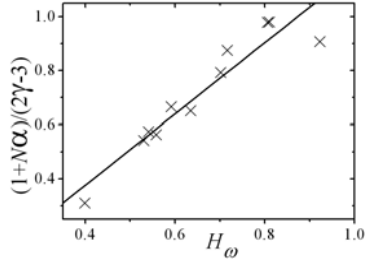

**Fig.2.** The empirically obtained (the forks) and the theoretical (the solid line) correlation between $H_\omega$ and $(1+N\alpha)/(2\gamma-3)$

It may be boring if we present also a figure to show the empirical correlation between $(1+N\alpha)/(2\gamma-3)$ and $H_s$. Because the correlation only shows a pure mathematic relation, the two figures must be almost the same no matter the physical quantities are different. However, the correlations between the two SPL parameters, $\alpha$ and $\gamma$, may be different for NNTW distribution, $P(\omega)$, and normalized act degree distribution, $P(s)$. For $P(s)$, from the mentioned Zhang-Chang model analysis, one knows that $\gamma=1+T/[(T-1)(1-p)]$ and $\alpha = p/[N(1-p)]$, thus he can easily obtain that $\gamma \approx N\alpha + 2$. The empirical SPL parameters of $P(s)$ should be in an agreement with this since Refs. [6-7] already showed the model relevance for the act degree evolution. However, nobody knows if the model also can be used for an explanation on the evolution of cooperation sharing or competition gain (the node weight). So, it may be interesting to show the empirical SPL parameter relation of NNTW distribution to see if it also obeys $\gamma \approx N\alpha + 2$. In fig. 3 the circles show the $N\alpha+2$ function of the empirical NNTW distribution SPL parameter $\gamma$ for 9 mentioned collaboration/competition networks. They are very nicely fitted by a linear line with the unit slope and a correlation coefficient 0.998. The other two mentioned systems (the 2004 Athens Olympic Game and the university matriculation network) show exceptionally high values of $N\alpha+2$. If we draw all the 11 empirical data, the figure will be something unshapely shown, but the linear fitting results will be exactly the same. For a comparison with the aforementioned results reported in Ref. [10], $\gamma \propto \alpha^{0.32}$, the forks in Fig. 3 show this function with exactly the same empirical data. It is clear that, although the qualitative tendencies, which the two data sets show, are in a agreement, the quantitative conclusion $\gamma \approx N\alpha + 2$ is much more accurate than the other one, $\gamma \propto \alpha^{0.32}$.

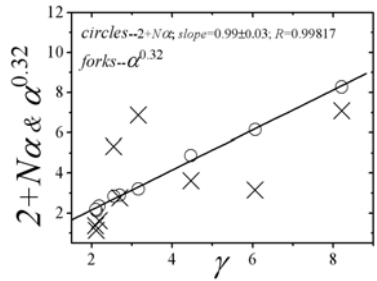

**Fig. 3.** Empirical data showing the $\alpha$ - $\gamma$ correlations suggested by Zhang-Chang model (circles) or Ref. [10] (forks), respectively.

How uneven is the distribution of the cooperation sharing or competition gain or the distribution of the cooperation/competition partner number in a cooperation/competition system? Why the distributions are very uneven in some systems and are quite even in some other systems? These are very important and interesting questions in many research fields. In this letter we show that the heterogeneity index, *H*, which was proposed in Ref. [16], can be used to describe the disparity of a distribution. The function between *H* and the SPL distribution parameters, *α* and *γ*, can be analytically deduced. Also, very surprisingly, we find that the analytic conclusion, $\gamma \approx N\alpha + 2$, which can be deduced by Zhang-Chang model, is in a very good agreement with the empirical investigation results about the distribution of the cooperation sharing or competition gain. This suggests that the model, in addition to describe the evolution of cooperation/competition partnership, may also be used for a description on the evolution of the cooperation sharing or competition gain. Therefore, it is possible to develop a new and general cooperation/competition dynamic model based on Zhang-Chang model that may answer the above questions. This will be reported in our future papers.